\documentclass[12pt,a4paper]{article}
\usepackage[latin1]{inputenc}
\usepackage{graphicx,psfrag}
\usepackage{amsmath,amssymb,amsfonts,amsthm}
\usepackage{multirow}
\def\n{\nonumber}
\def\be{\begin{equation}}
\def\ee{\end{equation}}
 
\theoremstyle{plain}

\theoremstyle{definition}

\title{The Deng algorithm in higher dimensions}
\author{Y Nyonyi\footnote{yusuf@aims.ac.za}, S D Maharaj\footnote{maharaj@ukzn.ac.za} \mbox{} and K S Govinder\footnote{govinder@ukzn.ac.za} \\ Astrophysics and Cosmology Research Unit\\School of Mathematics, Statistics and Computer Science\\University of KwaZulu-Natal\\ Private Bag X54001\\  Durban 4000, South Africa.}
\date{}
\begin{document}
\maketitle

\section*{Abstract}
We extend an algorithm of Deng in spherically symmetric spacetimes to higher dimensions. We show that it is possible to integrate the generalised condition of pressure isotropy and generate exact solutions to the Einstein field equations for a shear-free cosmological model with heat flow in higher dimensions. Three new metrics are identified which contain results of four dimensions as special cases. We show graphically that the matter variables are well behaved and the speed of sound is causal.

\section{Introduction}

Spherically symmetric gravitating models with heat flow, in the absence of shear, are important in the study of various cosmological processes and the evolution of relativistic astrophysical bodies. For a variety of applications in the presence of inhomogeneity see Krasinski $\cite{k8}$. Heat flow models are also important in analysing gravitational collapse and relativistic stellar processes. Astrophysical studies in which heat flow is important include the shear-free models of Wagh $\textit{et al}$ $\cite{wggmmm}$, Maharaj and Govender $\cite{mg}$, Misthry $\textit{et al}$ $\cite{mml}$ and Herrera $\textit{et al}$ $\cite{hdo}$. By studying shear-free models, we avail ourselves with a rather simpler avenue where we only need to provide solutions to the generalised condition of pressure isotropy containing two metric functions. A complete study of shear-free heat conducting fluids with charge was completed by Nyonyi $\textit{et al}$ $\cite{nmg1}$ using Lie's group theoretic approach applied to differential equations. Shearing models where heat flow is significant have been recently studied by Thirukkhanesh $\textit{et al}$ $\cite{trm1}$ for radiating spherically symmetric spheres. It turns out that the resulting nonlinear equations with shear are much more difficult to analyse.

A generic method of obtaining new solutions to the Einstein field equations with heat flow was provided by Deng $\cite{d1}$. Using this general method we can regain existing results and obtain new classes of solutions. Nyonyi $\textit{et al}$ $\cite{nmg1}$, Ivanov $\cite{ibv}$ and Msomi $\textit{et al}$ $\cite{mgm2}$ have obtained new solutions using the Lie group theoretic approach and other methods, by solving the underlying pressure isotropy condition. These investigations are applicable to four dimensions. Extensions to higher dimensions have also been considered by many authors because of physical requirements, for example, Bhui $\textit{et al}$ $\cite{bcb1}$ showed the absence of horizons in nonadiabatic gravitational collapse. Studies of this type motivated the Lie symmetry analysis of heat conducting fluids by Msomi $\textit{et al}$ $\cite{mgm3}$ in dimensions greater that four. In the present treatment, we extend the Deng $\cite{d1}$ algorithm to higher dimensions and show that new results are possible.

\section{The model} \label{nmodel}
We consider the line element of a shear-free, spherically symmetric $(n+2)$--dimensional manifold in the form
\be \label{nlineelement}  \mathrm{d}s^2 = -D^{2}\mathrm{d}t^2 + \frac{1}{V^{2}}\left(\mathrm{d}r^{2} + r^{2}\mathrm{d}X^{2}_{n} \right) \ee
where $n\geq 2$. The gravitational potential components $D$ and $V$ are functions of $r$ and $t$ with 
\be X^{2}_{n} = \mathrm{d}\theta^{2}_{1} + \sin^{2}\theta_{1}\mathrm{d}\theta^{2}_{2} +\dots +\sin^{2}\theta_{1}\sin^{2}\theta_{2}\dots \sin^{2}\theta_{n-1}\mathrm{d}\theta^{2}_{n} \ee
For a heat conducting fluid, the energy momentum tensor is given by
\be \label{EMomT} T_{ab} = (\rho + p) U_{a} U_{b} + pg_{ab} + q_{a} U_{b} + q_{b} U_{a} \ee 
where $\rho$ is the energy density, $p$ is the kinetic pressure, $\textbf{q}$ is the heat flux tensor and $\textbf{U}$ is a timelike $(n+2)$--velocity vector. For a comoving observer we have $U^{a} = \left(\frac{1}{D},0,0,\dots, 0\right)$ and $q^{a} = \left(0,q,0,\dots, 0\right)$.

Utilizing $\eqref{nlineelement}$--$\eqref{EMomT}$, we obtain the Einstein field equations
\begin{subequations} \label{nEINcomps}
\begin{align}
\rho &= \frac{n(n+1)V^{2}_{t}}{2D^{2}V^{2}} - \frac{n(n+1)VV^{2}_{r}}{2} + nVV_{rr} +\frac{n^{2}VV_{r}}{r} \label{nEINcomp1} \\ \n \\
p &= -\dfrac{nD_{r}VV_{r}}{D} + \frac{nD_{r}V^{2}}{rD}+ \frac{n(n-1)V_{r}^{2}}{2} - \frac{n(n-1)VV_{r}}{r} \n \\  
 & \quad + \frac{nV_{tt}}{D^{2}V} - \frac{n(n+3)V_{t}^{2}}{2D^{2}V^{2}} - \frac{nD_{t}V_{t}}{D^{3}V} \label{nEINcomp2} \\ \n \\
p &= \dfrac{D_{rr}V^{2}}{D} - (n-1)VV_{rr} + \frac{n(n-1)V_{r}^{2}}{2}  + \frac{(n-1)D_{r}V^{2}}{rD} -  \frac{(n-1)^{2}VV_{r}}{r} \n \\ 
&\quad - \frac{(n-2)D_{r}VV_{r}}{D}  + \frac{nV_{tt}}{D^{2}V} - \frac{n(n+3)V_{t}^{2}}{2D^{2}V^{2}} - \frac{nD_{t}V_{t}}{D^{3}V} \label{nEINcomp3} \\ \n \\
q &= -\frac{nVV_{tr}}{D} + \frac{nV_{r}V_{t}}{D} +\frac{nD_{r}VV_{t}}{D^{2}} \label{nEINcomp4}  
\end{align} 
\end{subequations} 
which are consistent with the derivation of Bhui $\textit{et al}$ $\cite{bcb1}$. Equations $\eqref{nEINcomp2}$ and $\eqref{nEINcomp3}$, together with the transformation $u =r^{2}$, give the pressure isotropy condition
\be \label{nisotropy}  VD_{uu} +2D_{u}V_{u} - (n-1)DV_{uu} = 0  \ee 
which is the master equation for the gravitating fluid in $(n+2)$--dimensions. 

Deng $\cite{d1}$ provided a general recipe for generating a series of solutions of the isotropy condition $\eqref{nisotropy}$ for $n=2$. This technique may be extended to the master equation $\eqref{nisotropy}$. Note that the isotropy condition is an ordinary differential equation in $u$ (since no time derivatives appear) which may be reduced to a simpler differential equation in $D$ if $V$ is known and vice versa. In this technique, elementary forms of either $V$ or $D$ are chosen that are in turn substituted into the equation to be solved so as to obtain the form of the remaining term. Below we give a brief outline of the method.
\begin{enumerate}
\item Take a simple form of $V$, say $V = V_{1}$, and substitute it into the master equation to find the most general solution of $D$, say $D = D_{1}$. The pair  $V = V_{1}$ and  $D = D_{1}$ provides the first class of solutions to $\eqref{nisotropy}$.
\item Take $D = D_{1}$ and substitute it into the master equation. This gives an equation in $V$ with $V = V_{1}$ already a solution. We are now in a position to obtain a second solution  $V = V_{2}$ linearly independent of  $V_{1}$. The linear combination $V_{3} = aV_{2} + bV_{1}$ gives the general solution that satisfies the master equation. The pair $V = V_{3}$ and  $D = D_{1}$ is the second class of solutions to $\eqref{nisotropy}$.
\item Take $V = V_{3}$ and substitute it into the master equation. We obtain an equation in $D$ with $D=D_{1}$ already a solution. We are then in a position to obtain $D=D_{2}$ in the same way we obtained $V_{2}$. The pair $V = V_{3}$ and  $D = cD_{1}+ dD_{2}$ is the third class of solutions to $\eqref{nisotropy}$.
\item Repeat the above process to obtain an infinite sequence of solutions.
\end{enumerate} 
It is important to note that, in principle, this is a non-terminating process for obtaining solutions, and an infinite number of solutions can be listed. The difficulty arises in obtaining subsequent solutions in the process because the integration may become more complicated. However, the algorithm proves to be a powerful mechanism for generating new solutions.

\section{Results} \label{result}
We start with a simple case
\be \label{d1} D_{1} = 1 \ee
Equation $\eqref{nisotropy}$ reduces to
\be \label{vprime2} V_{uu} = 0 \ee
This equation can be solved directly to obtain
\be  \label{v1} V_{1} = au + b \ee
where $a$ and $b$ are arbitrary functions of $t$. The pair of equations $\eqref{d1}$ and $\eqref{v1}$ gives the first class of solutions. 
\be \label{nel1}\mathrm{d}s^2 = -\mathrm{d}t^2 + \frac{1}{(au+b)^2}\left(\mathrm{d}r^{2} + r^{2}\mathrm{d}X^{2}_{n} \right) \ee
Observe that the dimension $n$ is absent in the metric $\eqref{nel1}$. In this class of solution the potentials are independent of the dimensionality; however the spatial
volume depends on $n$ and the matter variables are affected through the field equations. When $n=2$ we regain the results of Bergmann $\cite{b1}$. 

On substituting $\eqref{v1}$ back into $\eqref{nisotropy}$ we obtain
\be \label{dd2}(au+b)D_{uu} + 2aD_{u} = 0 \ee
The general solution to equation $\eqref{dd2}$ is
\be \label{d2} D_{2} = \dfrac{cu+d}{au+b} \ee
with $c$ and $d$ arbitrary functions of $t$. The pair of equations $\eqref{v1}$ and $\eqref{d2}$ gives the second class of solutions
\be \label{nel2}\mathrm{d}s^2 = -\left(\dfrac{cu+d}{au+b}\right)^{2}\mathrm{d}t^2 + \frac{1}{(au+b)^{2}}\left(\mathrm{d}r^{2} + r^{2}\mathrm{d}X^{2}_{n} \right) \ee
Again, we observe that the dimension $n$ does not appear explicitly in $\eqref{nel2}$. This means that the potentials in $\eqref{nel2}$ are independent of the dimension. When $n=2$, we regain the solutions obtained by Maiti $\cite{m1}$ and later generalised by Modak $\cite{m2}$ and Sanyal and Ray $\cite{sr}$. We make the general point that for a linear form of $V$, the parameter $n$ does not appear in equation $\eqref{nisotropy}$. Thus all solutions with a linear form for $V$ do not contain the dimension $n$, thereby leading to the metrics $\eqref{nel1}$ and $\eqref{nel2}$.

Now, substituting $\eqref{d2}$ into $\eqref{nisotropy}$, we obtain
\be \label{v22} V_{uu} - \dfrac{2}{n-1} \left( \dfrac{(bc-ad)/(au+b)^2}{(cu+d)/(au+b)}\right) V_{u} + \dfrac{2}{n-1} \left( \dfrac{a(bc-ad)/(au+b)^3}{(cu+d)/(au+b)}\right) V = 0 \ee
We require two independent solutions $V_{1}$ and $V_{2}$ of $\eqref{v22}$. Note that $V_1$ in $\eqref{v1}$ is a solution to $\eqref{v22}$. We propose the second solution to $\eqref{v22}$ to be given by
\be \label{v221} V_{2} = \alpha(u,t) V_{1} \ee
where the function $\alpha(u,t)$ has to be found explicitly. On substituting $\eqref{v221}$ into $\eqref{v22}$, we obtain
\be \label{alpha1} \alpha_{uu} + 2\left[\dfrac{a}{au+b} - \dfrac{1}{n-1} \left( \dfrac{(bc-ad)/(au+b)^2}{(cu+d)/(au+b)}\right) \right]\alpha_{u} = 0 \ee
On integrating $\eqref{alpha1}$, we obtain $\alpha$ expressed as 
\be \label{alpha3}\alpha =\int^{u} e \left( \dfrac{cs+d}{(as+b)^n} \right)^{2/(n-1)} \mathrm{d}s \ee
Consequently, the second solution $V_{2}$ will depend on the dimension $n$. To evaluate the integral $\eqref{alpha3}$, we need to consider two cases: $ad = bc$ and $ad \neq bc$.

\subsection{Case I}
When $ad = bc$ we have
\begin{align} \label{alpha4}
  \alpha &= e \left( \frac{d}{b^{n}} \right)^{\frac{2}{n-1}} \int \left(1 + \dfrac{a}{b}u \right)^{-2} \mathrm{d}u \nonumber \\
         &= \dfrac{a^{2}gu+(abg-kb^{2})}{a(au+b)}
\end{align}         
where $g$ is an arbitrary function of $t$. Therefore $V_{2}$ becomes
\be \label{vstar}  V_{2} = agu+\left(bg-\frac{k}{a}b^{2}\right) \ee
This implies that $V_{2}$ is proportional to $V_{1}$ and is therefore not a second linearly independent solution. The case $ad = bc$ is degenerate.

\subsection{Case II}
When $ad \neq bc$, we have
\be \alpha = \frac{e}{ad-bc} \left(\frac{1-n}{n+1}\right) \left( \dfrac{cu+d}{au+b} \right)^{\frac{n+1}{n-1}} + g \ee
where $a$, $b$, $c$, $d$, $e$ and $g$ are arbitrary functions of $t$. Therefore the second solution $V_{2}$ becomes
\be  V_{2} = \left( \dfrac{e}{ad-bc}\left(\dfrac{1-n}{n+1}\right)\left( \dfrac{cu+d}{au+b} \right)^{\frac{n+1}{n-1}} + g \right)(au+b)\ee
And since the general solution to $\eqref{nisotropy}$ is linear combination $V_1$ and $V_2$, we obtain
\be \label{v3} V_{3} = \left( h(t)+ j(t)  \left( \dfrac{e}{ad-bc}\left(\dfrac{1-n}{n+1}\right)\left( \dfrac{cu+d}{au+b} \right)^{\frac{n+1}{n-1}} + g \right)\right) \left( au+b\right) 
\ee
where we have introduced for convenience $h(t)$ and $j(t)$. The third class of solutions is therefore given by $\eqref{d2}$ and $\eqref{v3}$ with metric
\begin{align} \label{nel3}\mathrm{d}s^2 &= -\left(\dfrac{cu+d}{au+b}\right)^{2} \mathrm{d}t^2 \nonumber \\
& \mbox{  }\quad + \left( \left[ h(t)+ j(t)  \left( \dfrac{e}{ad-bc}\left(\dfrac{1-n}{n+1}\right)\left( \dfrac{cu+d}{au+b}\right)^{\frac{n+1}{n-1}} + g \right)\right] \left( au+b\right) \right)^{-2} \nonumber \\
& \mbox{  }\quad \times \left(\mathrm{d}r^{2} + r^{2}\mathrm{d}X^{2}_{n} \right) \end{align}
This is a new class of solution and it is evident that it certainly depends on the dimension $n\geq 2$. Therefore we can conclude that the dimensionality of the problem does indeed affect the dynamics of the gravitational field with heat flow. The next class of solutions can be obtained by substituting $V_{3}$ into equation $\eqref{nisotropy}$ and then solve the resulting equation for $D_{3}$. This may be continued to obtain further new solutions. The integration process gets more complicated for further iterations.

We now consider the special case of four dimensions. When $e=1$ and $n=2$ the line element $\eqref{nel3}$ becomes
\begin{align} \label{nel4}\mathrm{d}s^2 &= -\left(\dfrac{cu+d}{au+b}\right)^{2}\mathrm{d}t^2 \nonumber \\
& \mbox{ }\quad + \left( \left[ h - j\left( \dfrac{1}{3(ad-bc)}\left( \dfrac{cu+d}{au+b}\right)^{3} + g \right)\right] \left( au+b\right) \right)^{-2} \nonumber \\ & \mbox{ }\quad \times \left(\mathrm{d}r^{2} + r^{2}(\mathrm{d} \theta^{2} + \sin^{2} \theta^{2}\mathrm{d} \phi^{2}) \right) \end{align}
We can rewrite $\eqref{nel4}$ in the equivalent form 
\begin{align} \label{nel5}\mathrm{d}s^2 &= -\left(\dfrac{cu+d}{au+b}\right)^{2} \mathrm{d}t^2 \nonumber \\
& \mbox{ } \quad +\left( (h+\kappa )\left( au+b \right) - \frac{j}{3a} \left( \frac{c^{2}}{a^{2}} + \frac{c}{a} \frac{au+b}{cu+d} +\left(\frac{cu+d}{au+b}\right)^{2} \right)\right)^{-2} \nonumber \\ & \mbox{ }\quad \times \left(\mathrm{d}r^{2} + r^{2}(\mathrm{d} \theta^{2} + \sin^{2} \theta^{2}\mathrm{d} \phi^{2}) \right)
\end{align}
where the function $\kappa$ is given by 
\be \label{kap2} \kappa = \left( g- \frac{c^{3}}{3a^{3}(ad-bc)} \right)\ee
When we set $\kappa = 0$ in $\eqref{nel5}$ we regain the result of Deng $\cite{d1}$. We interpret $\eqref{nel3}$ as the higher dimensional generalisation of the Deng model with heat flow.

\section{Example}
\begin{figure}[h!]
\begin{center}
\includegraphics[scale=0.5]{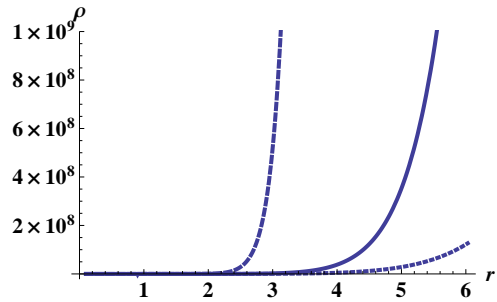}
\caption{Energy density $\rho$}
\label{engyden}
\end{center}
\end{figure}

\begin{figure}[h!]
\begin{center}
\includegraphics[scale=0.6]{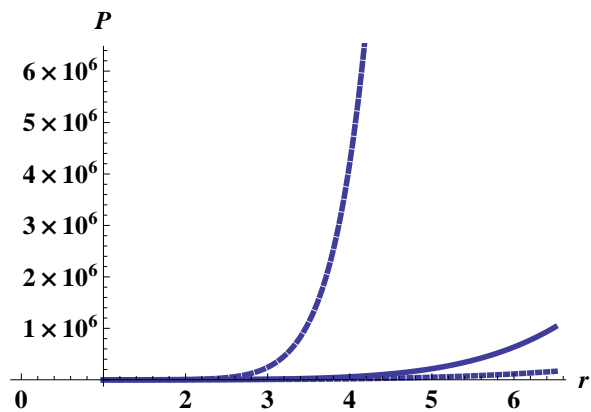}
\caption{Pressure}
\label{pressure}
\end{center}
\end{figure}

\begin{figure}[h!]
\begin{center}
\includegraphics[scale=0.6]{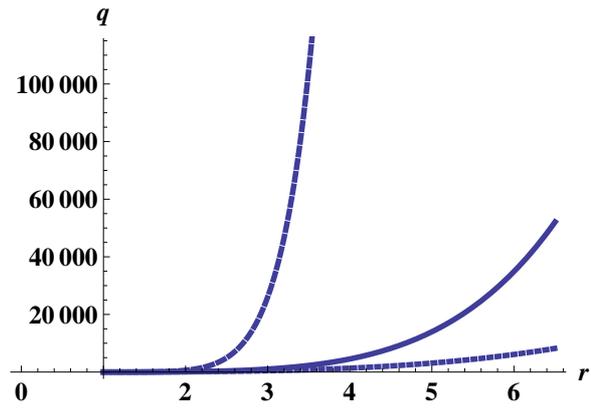}
\caption{Heat flow}
\label{heatf}
\end{center}
\end{figure}

\begin{figure}
\begin{center}
\includegraphics[scale=0.6]{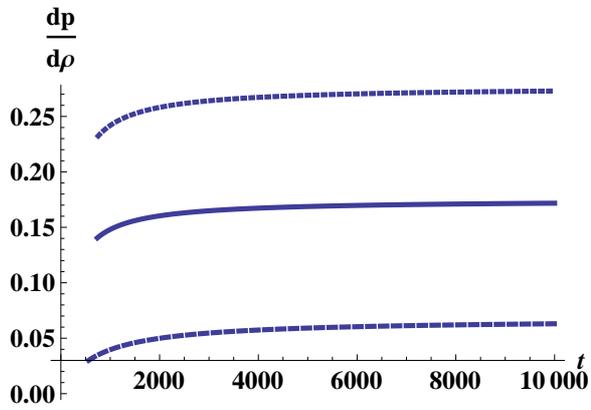}
\caption{Speed of sound}
\label{soundspd}
\end{center}
\end{figure}

\begin{figure}[h!]
\begin{center}
\includegraphics[scale=0.6]{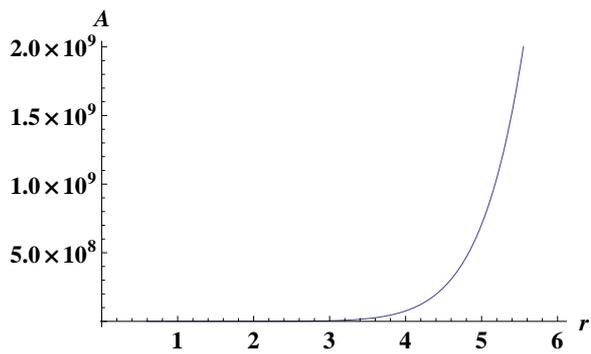}
\caption{Weak energy condition}
\label{weakcon}
\end{center}
\end{figure}

\begin{figure}[h!]
\begin{center}
\includegraphics[scale=0.6]{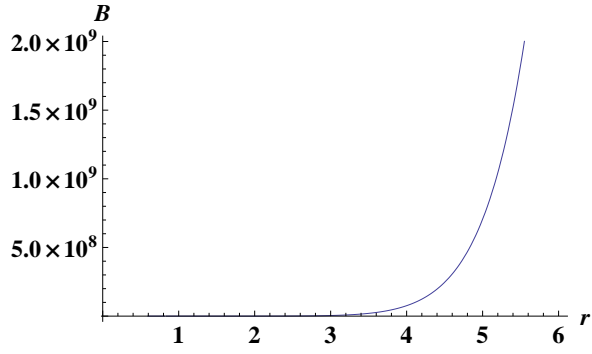}
\caption{Dominant energy condition}
\label{domcon}
\end{center}
\end{figure}

\begin{figure}[h!]
\begin{center}
\includegraphics[scale=0.6]{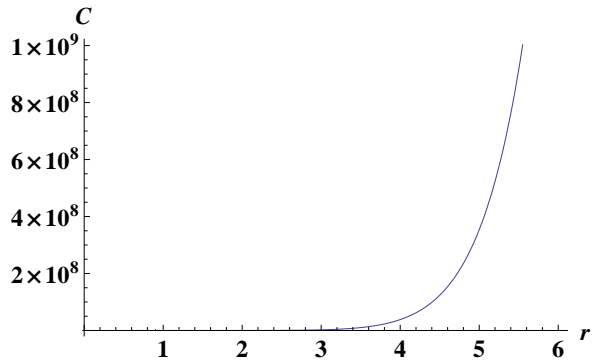}
\caption{Strong energy condition}
\label{strong}
\end{center}
\end{figure}

We illustrate the validity of our solutions by considering a simple example with physically viable conditions. For the line element $\eqref{nel3}$, we make the simple choice: $a=d=0$, $b=c=1$, $h+jg =1$ and $e=j=1$. This gives the simplified forms of the potentials
\be D = r^2, \qquad V= 1 + \left(\frac{1-n}{n+1}\right) t\mbox{ } r^{(2(n+1)/(n-1))} \ee
Even with this simple example, a qualitative analysis of the matter variables and energy conditions for the interior matter distribution is arduous. We therefore generate graphical plots on a constant timelike hypersurface to illustrate the validity of our solutions using this example. Figures $\ref{engyden}$--$\ref{heatf}$ are the plots for the energy density $\rho$, the  pressure $p$ and the heat flow $q$ for three different dimensions: $n=2$ (dashed line), $n=3$ (solid line) and $n=4$ (dotted line). It is clearly evident that the matter variables are positive and they decrease with increase in dimension. This is due to the fact that an increase in dimension translates to an increase in the number of degrees of freedom leading to a decrease in the mass per unit volume of the gravitating fluid. In Figure $\ref{soundspd}$ we have plotted the speed of sound. From Figure $\ref{soundspd}$, we observe that causality is not violated for the dimensions $n=2$, $3$ and $4$. In Figures $\ref{weakcon}$--$\ref{strong}$ we have plotted the quantities $A =\rho -p +\Delta $, $B= \rho -3p + \Delta$ and $C= 2p + \Delta$, where $\Delta= \sqrt{(p+q)^2 - 4q^2}$ for $n=3$. We observe that $A$, $B$ and $C$ are positive; hence the weak, dominant and strong energy conditions are satisfied. Therefore the matter distribution for this example is physically reasonable.

\section{Discussion}
We obtained new generalised classes of exact solutions to the Einstein field equations for a neutral relativistic fluid in the presence of heat flow in a higher dimensional manifold. We found new solutions to the coupled Einstein system by solving the higher dimensional pressure isotropy condition which is a second order nonlinear differential equation. We solved the master equation by making use of the Deng algorithm $\cite{d1}$ and obtained three new metrics. The first metric $\eqref{nel1}$ generalises the Bergmann $\cite{b1}$ line element. The second metric $\eqref{nel2}$ generalises the Maiti $\cite{m1}$, Modak $\cite{m2}$ and Sanyal and Ray $\cite{sr}$ line elements. It is remarkable that the potentials in $\eqref{nel1}$ and $\eqref{nel2}$ are independent of the dimension. The third metric $\eqref{nel3}$ depends on the dimension $n$ and generalises the Deng $\cite{d1}$ line element. We conclude that the dimension of the spacetime affects the dynamics of the heat conducting gravitating fluid. We briefly studied the physical features by graphically plotting the matter variables. The energy conditions are found to be positive and causality is not violated.

\section*{Acknowledgements}
YN, KSG and SDM wish to thank the National Research Foundation and the University of KwaZulu-Natal for financial support. SDM further acknowledges that this research is supported by the South African Research Chair Initiative of the Department of Science and Technology

 \end{document}